\documentclass[aps,prb,twocolumn,superscriptaddress,showpacs,floatfix]{revtex4}
\usepackage{graphicx}
\usepackage{dcolumn}
\usepackage{bm}
\usepackage{color}

\newcommand{\ef}{E_\mathrm{F}}

\bibstyle{nature}

\begin{document}
\title{Proximity induced topological state in graphene}

\author{Igor Popov}
\author{Mauro Mantega}
\author{Awadhesh Narayan}
\email{narayaa@tcd.ie}
\author{Stefano Sanvito}

\affiliation{School of Physics, AMBER and CRANN, Trinity College, Dublin 2, Ireland}

\date{\today}


\begin{abstract}
The appearance of topologically protected states at the surface of an ordinary insulator is a rare occurrence and to date only a 
handful of materials are known for having this property. An intriguing question concerns the possibility of forming  topologically 
protected interfaces between different materials. Here we propose that a topological phase can be transferred to graphene by 
proximity with the three-dimensional topological insulator Bi$_2$Se$_3$. By using density functional and transport theory we 
prove that, at the verge of the chemical bond formation, a hybrid state forms at the graphene/Bi$_2$Se$_3$ interface. The state 
has Dirac-cone-like dispersion at the $\Gamma$ point and a well-defined helical spin-texture, indicating its topologically protected 
nature. This demonstrates that proximity can transfer the topological phase from Bi$_2$Se$_3$ to graphene.
\end{abstract}

\maketitle

Topological insulators (TIs) are a recently discovered class of materials presenting an electronic band-gap in the bulk and 
metallic edge states at their surfaces~\cite{TIreview1, TIreview2, firstNatPaperOnBise}. The peculiarity is that the edge states 
are protected against electron scattering to impurities, so that they can act as perfectly ballistic conductors~\cite{TIreview1,Park}. 
To date only a handful of TIs have been synthesized and high-throughput materials screening has indicated that only few more 
may be fabricated by straining existing inorganic compounds~\cite{Curta}. Intriguingly, with the only exception of CdTe/HgTe-quantum 
wells~\cite{Mole}, there are no reports of two-dimensional (2D) TIs. Thus at the moment we have at hand only three-dimensional (3D) 
TIs with 2D edge states. Yet, one may wonder whether we have exhausted all the possibilities for creating useful topologically 
protected states in materials.

A particularly intriguing prospect is that of using the interaction between different materials to create hybrid interfaces with topological
properties. For instance depositing normal semiconductors on top of 3D topological insulators may result in a structure that under 
certain conditions exhibits topologically protected interface states \cite{Cesare}.  An even more attractive prospect is that of using 
that protocol for transferring topologically protected states to graphene~\cite{GrapheneReview1,GrapheneReview2}. Since 
graphene-based transistors have been already demonstrated~\cite{GrapTrans}, one could then speculate on having graphene logic 
elements connected by topological-graphene interconnects, i.e. on realizing an all graphene high-performance logic circuitry. A major 
advantage of such strategy is its fully compatible with 2D patterning.
\begin{figure}
\includegraphics[width=\columnwidth]{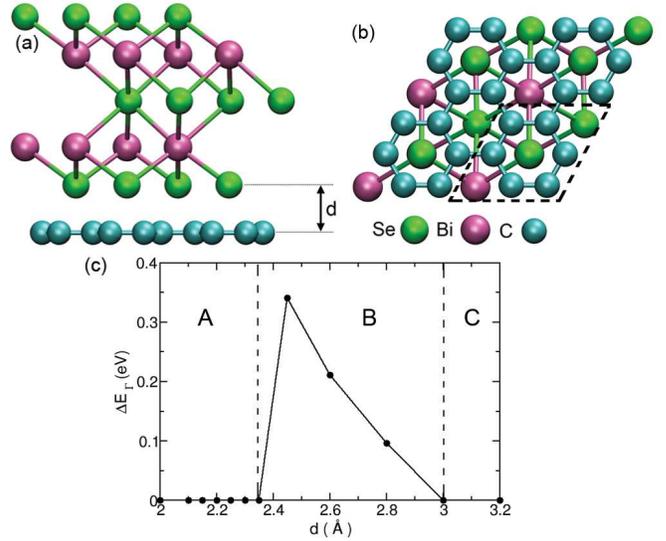}
\caption{\label{fig1} (color online) Side (a) and top view (b) of the graphene/Bi${_2}$Se${_3}$ interface. 
The graphene-Bi${_2}$Se${_3}$ separation is $d$. In panel (c) we report the graphene electronic band gap (see text) as a function 
of $d$.}
\end{figure}

Several proposals have been already brought forward for making graphene topological~\cite{TIpropsGraphene1}. Indeed
one of the first TIs models was based on a staggered hexagonal lattice with helicity-dependent complex hopping parameter, 
simulating spin-orbit interaction~\cite{KaneMele}. However, since spin-orbit coupling in graphene is tiny, a topological phase 
may be induced only by strongly perturbing the graphene electrostatic potential, for instance by adsorbing heavy 
ions on top of the sheet~\cite{HeavyAdatomsOnGraphene}. Importantly, although theoretically sound, such proposal requires 
ultra-accurate fabrication precision and appears rather challenging in practice. Here we suggest a completely different approach: 
we introduce topologically protected states in graphene by proximity with a lattice-commensurate 3D TI. This happens at the 
graphene/Bi${_2}$Se${_3}$ interface, a composite which was synthesized about two years 
ago~\cite{Nanolett_BiSeGraphene,APL_BiSeGraphene, PRL_GrapheneSb2Te3}, but whose electronic structure still remains 
unclear.

\section{Computational methods}

Calculations are performed by density functional theory (DFT) as implemented in the {\sc VASP} code~\cite{VASP1,VASP2}. 
We use the Perdew-Burke-Ernzerhof form of the generalized gradient approximation~\cite{PBE} and the core electrons 
are described by projector-augmented-wave pseudopotentials~\cite{PAW}. The $k$-space integration spans a 
$11\times 11 \times 1$ Monkhorst-Pack mesh in the irreducible Brillouin zone and the plane waves cutoff is 400~eV. 

The geometry of the structure investigated is shown in Fig.~\ref{fig1}. We consider a Bi${_2}$Se${_3}$ slab containing 
three quintuple layers (QLs), for which the tensile stress is minimal among the experimentally investigated Bi${_2}$Se${_3}$/graphene 
composites~\cite{Nanolett_BiSeGraphene}. The Bi${_2}$Se${_3}$ unit cell is commensurate with three graphene 
unit cells, hence the elementary unit cell of the composite contains an entire carbon ring. The contacting Se atom is placed at the 
graphene hollow site (in the center of the ring). The in-plane lattice parameter is 4.26~\AA, which is only 2.3\% larger than the 
lattice parameter of bulk Bi${_2}$Se${_3}$~\cite{BiSeGeom}; the one perpendicular to the interface is instead 40~\AA\ (there is 
a vacuum region $>$10~\AA\ between cells periodic replica). We have also investigated a second geometry 
where the carbon atoms are on top of Se. This, however, is not energetically favorable and it has not been considered in the rest 
of the paper. Interestingly both interface structures present rather similar trends in the electronic structure properties.

\begin{figure*}
\includegraphics[width=\textwidth]{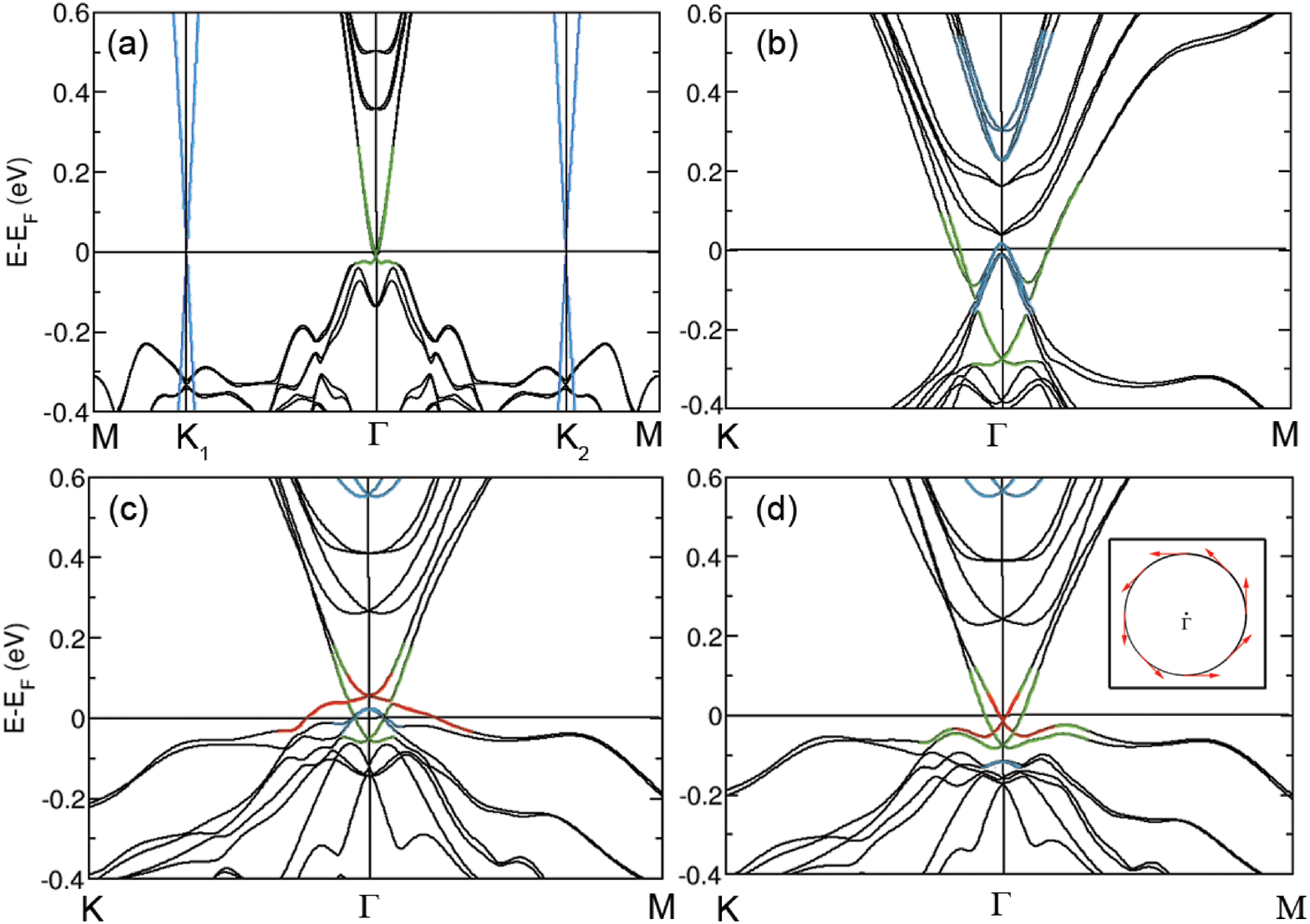}
\caption{\label{fig2} (color online) Evolution of band-structure of the graphene/Bi$_2$Se$_3$ composite as a function of the separation
$d$ between the two constituents. In panels (a), (b), (c) and (d) we present the band-structure for $d=3.0$~\AA, 2.6~\AA, 2.3~\AA\ and 
2.2~\AA\ respectively. Black and green bands are bulk and surface states of Bi${_2}$Se${_3}$, blue bands are graphene bands, while 
the red ones represent hybrid states. The inset in the panel (d) illustrates the spin-texture of the mixed state at 0.05~eV above $\ef$. 
Note the different $k$-point samplings for $d = 3.0$~\AA.}
\end{figure*}

\section{Results and discussion}

Let us begin by investigating the evolution of the graphene band-gap with the graphene/Bi$_2$Se$_3$ 
distance, $d$. We assign an electronic band to a given material by projecting the energy and $k$-dependent wave function 
onto spherical harmonics centered around particular atoms~\cite{BandContrib}. We define the graphene electronic 
band-gap from those bands located near the charge neutral point of free-standing graphene and having dominant C character. 
In Fig.~\ref{fig1}(c) one can identify three different regions. For $d>3$~\AA\ [region C] graphene has no band-gap. This is 
expected since for large separations the interaction is weak and the band-structure of the composite is the superposition of 
those of the constituents. As such graphene remains a zero-gap semiconductor. Region B is characterized 
by the opening of the graphene band-gap. The gap increases monotonically from $d=3$~\AA\ and it reaches a maximum 
(0.34~eV) for $d=2.45$~\AA. A further reduction in $d$ (region A) closes the gap, which remains close up to $d=2$~\AA.

\begin{figure}
\includegraphics[width=\columnwidth]{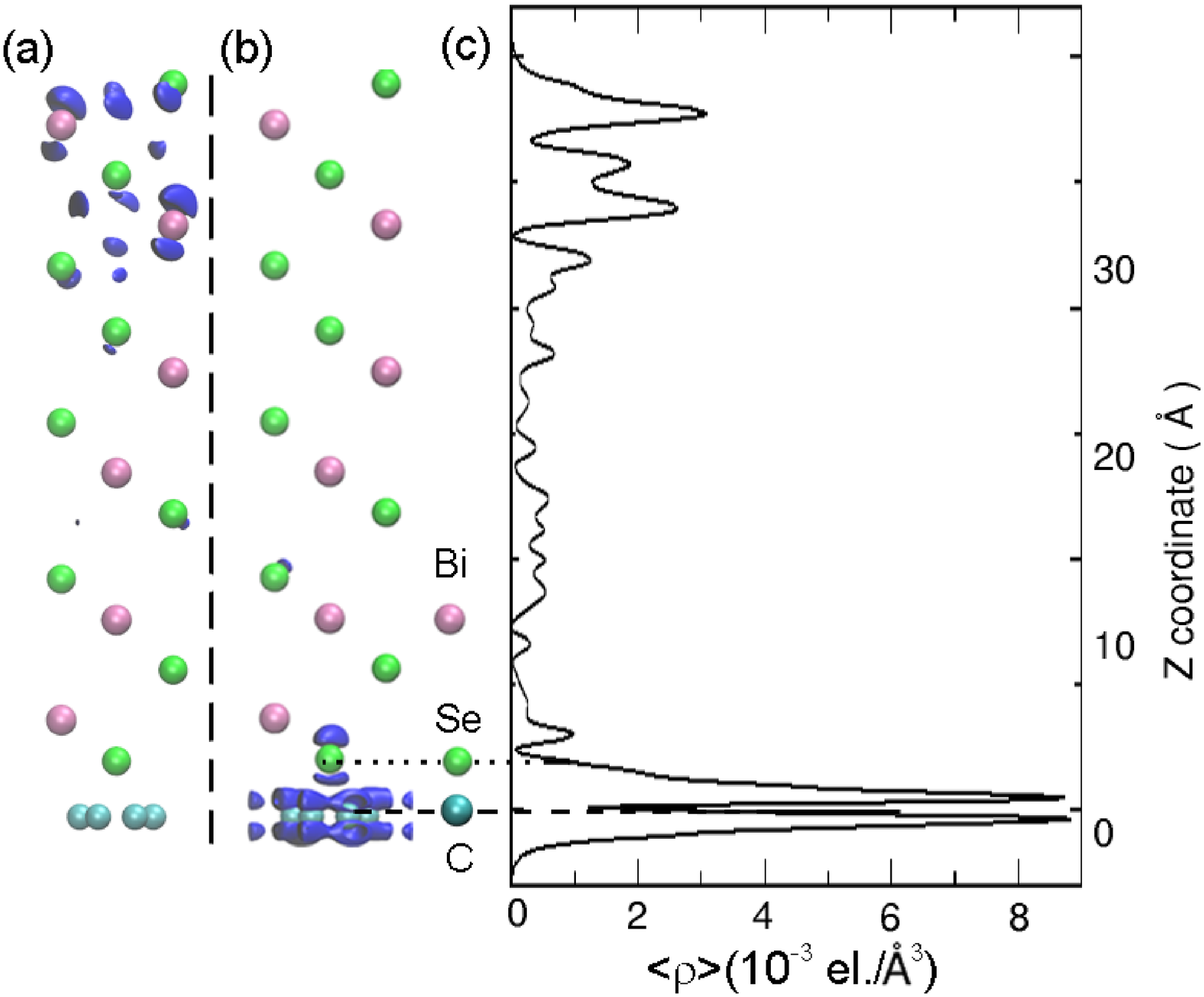}
\caption{\label{fig3} (color online) Charge density associated to the Bi${_2}$Se${_3}$ surface opposite to the graphene/Bi${_2}$Se${_3}$ 
interface (a) and the mixed interface state (b) obtained for $d = 2.2$~\AA\ at $\Gamma$. Panel (c) shows the sum of the two charge 
densities averaged over a plane parallel to the interface.}
\end{figure}

Next we analyze in Fig.~\ref{fig2} the nature of the graphene bands around the Fermi level, $\ef$, as a function of $d$. 
For $d > 3$~\AA\ [Fig.~\ref{fig2}(a)] the composite features two superimposed band-structures corresponding to 
those of graphene and Bi${_2}$Se${_3}$, respectively. At such large separation there is no wave function overlap between 
graphene and Bi${_2}$Se${_3}$, leaving the two materials electronically decoupled. The graphene's linearly dispersive bands 
(in blue in Fig.~\ref{fig2}) at each of the valleys (Dirac K-points) are 2-fold spin-degenerate, with the $\pi$ ($E<\ef$) and 
$\pi^*$ ($E>\ef$) bands just touch each other. The Bi${_2}$Se${_3}$ surface states (green bands in Fig.~\ref{fig2}) cross 
$E_\mathrm{F}$ at the $\Gamma$ point.

Decreasing $d$ below 3~\AA\ causes a band-gap opening between the $\pi$ and $\pi^*$ bands [see Fig.~\ref{fig2}(b), $d=2.6$~\AA]. 
Now the graphene valleys are placed together with the Bi${_2}$Se${_3}$ surface states around the $\Gamma$ point
due to the bands folding in the supercell structure. As the graphene electronic gap increases further upon a reduction 
of $d$ [see $d=2.4$~\AA, Fig.~\ref{fig2}(c)], the $\pi^*$ cone lifts up in energy but the tip of the $\pi$ one remains pinned 
at $E_\mathrm{F}$. The topologically protected surface states of Bi${_2}$Se${_3}$ (one per surface) are positioned in the vicinity 
of the tip of the $\pi$ cone. For separations $d > 2.6$~\AA\ these surface states form a doubly-degenerate state since the two surfaces 
are equivalent for the unperturbed TI slab. For distances $d \leq 2.6$~\AA\ the symmetry of the TI slab breaks due to the vicinity of the 
graphene layer and the degeneracy of the surface states is lifted. The TI surface state in contact with graphene moves up in energy at 
$\Gamma$ by about 0.11~eV for $d=2.3$~\AA. Importantly, the graphene states that are pinned at $\ef$ start to couple with the surface 
state and, for $d\leq2.3$~\AA, the 4-fold degeneracy of the $\pi$ cone is lifted. Here the pure graphene bands forming the $\pi$ cone are 
pushed down in energy and only the mixed graphene/Bi${_2}$Se${_3}$ band [red in Fig.~\ref{fig2}(d)] and the Bi${_2}$Se${_3}$ surface 
state at the opposite side of the interface (placed directly beneath the mixed band at $\Gamma$) cross $E_\mathrm{F}$.

Intriguingly, such newly formed mixed graphene/Bi${_2}$Se${_3}$ band presents a helical spin-texture, demonstrated in the inset of 
Fig.~\ref{fig2}(d). This is a sufficient condition for disabling back-scattering of charge carriers~\cite{TIreview1}, and it is not 
the case in a pristine graphene sheet. In fact, defects in graphene allow hopping of charge carriers between two valleys, which causes 
back-scattering due to their opposite winding numbers. In contrast, hopping is impossible in systems with only one valley and a helical 
spin-texture, which is the case for graphene/Bi$_2$Se$_3$. Note, however, that the existence of the helical spin texture 
and the lack of back-scattering do not necessarily mean topological protection of the material's electronic state~\cite{spinTexture}.

The supercell structure causes the folding of the second Brillouin zone (BZ) of primitive graphene into the first and consequently the 
migration of the graphene valleys from K to $\Gamma$. Thus two 4-fold degenerate cones touching at $E_\mathrm{F}$ are formed. 
The bands from the two valleys have opposite topological charges, which causes their mutual 
annihilation~\cite{Vozmediano_GuageFieldsGraphene}. This manifests itself in the opening in graphene of a band-gap 
[see Fig.~\ref{fig2}(b) and (c)], much larger than that estimated  for thallium adatoms deposition \cite{HeavyAdatomsOnGraphene}.

After having determined the emergence of a mixed graphene/Bi$_2$Se$_3$ band, we now analyze in detail its electronic 
properties. The electron density integrated over a narrow energy region around $\ef$ and projected over the mixed state is 
shown in Fig.~\ref{fig3}(b). This is clearly localized over graphene and, to a smaller degree, over the Se atoms in contact to 
graphene. Since such state presents a dominating C-$p_z$ and Se-$p_z$ orbital contribution and it is delocalized in the plane 
of the interface, it presents $\pi$ conjugation. A more quantitative insight is obtained by plotting the charge density averaged 
over planes parallel to the interface [Fig.~\ref{fig3}(c)]. This shows that, while the contribution to the electron density originating 
from the bulk is small, a much larger portion is provided by the two surface states at both sides of the composite. At the free 
Bi$_2$Se$_3$ surface the surface state is distributed mainly over the first four atomic layers [Fig.~\ref{fig3}(a)]. In contrast, at 
the graphene/Bi$_2$Se$_3$ interface the electron density migrates from the TI to graphene. Notably this feature resembles 
closely the one reported for the interface between the normal metal Sb$_2$Se$_3$ and the TI 
Bi$_2$Se$_3$~\cite{Sb2Se3_Bi2Se3}. Also the behavior is similar to the topologization of ZnM (M=S, Se, Te) upon 
deposition on Bi$_2$Se$_3$~\cite{Cesare}.

Let us now spend a few words on the possibility of inducing a topological state in graphene due to its proximity and bonding to 
Bi$_2$Se$_3$. Firstly, we wish to point out that our results do not indicate that graphene converts into a 2D topological insulator 
upon its deposition on Bi$_2$Se$_3$, but simply that a topologically-protected hybrid state is formed. A 2D TI is an insulator in 
the bulk presenting topologically protected 1D states at the edges of a ribbon, as in the case of Bi thin films on Bi$_2$Te$_3$ 
surface \cite{Hirahara, Yang}. In contrast here the Fermi surface of graphene undergoes a transition from a zero-band gap 
semiconducting phase (region C), prone to gap opening due to defects and impurities, to a topologically protected metallic 
phase (region A) via an insulating phase (region B). A fundamental property of 3D TIs is the existence of an odd number of 
surface bands around $\ef$. Another property is that the topology of the surface states is such to connect the bulk valence 
band to the conduction one due to the parity inversion originated by the strong spin-orbit coupling. In the graphene/Bi$_2$Se$_3$ 
complex only one conical band is present at the Fermi level. Importantly this band belongs to the surface state of Bi$_2$Se$_3$ 
with considerable contribution of graphene around $\ef$, while the other parts of the state (in particular the ends that connect to 
the valance and the conduction bulk TI states) still fully belong to the TI. Thus the topologically protected surface TI state can be 
understood as a carrier of the induced graphene states, and the intrinsic topological protection of TI surface state provides 
the robustness to graphene as well. 

The topologically-protected hybrid state does not simply correspond to the penetration of the one of the Bi$_2$Se$_3$ edge
states into graphene. In that situation interaction between Bi$_2$Se$_3$ and graphene is not present. In contrast here the interaction is 
strong and, in fact, as $d$ decreases first it is responsible for the opening of a band-gap in graphene and then for creating the 
topologically-protected hybrid state. This behaviour is very similar to that of the Sb$_2$Se$_3$/Bi$_2$Se$_3$~\cite{Sb2Se3_Bi2Se3} 
and the ZnM/Bi$_2$Se$_3$ (M=S, Se, Te)~\cite{Cesare} interfaces, in which a topological state is transferred to the normal metal 
because of proximity.

\begin{figure*}
\includegraphics[width=\textwidth]{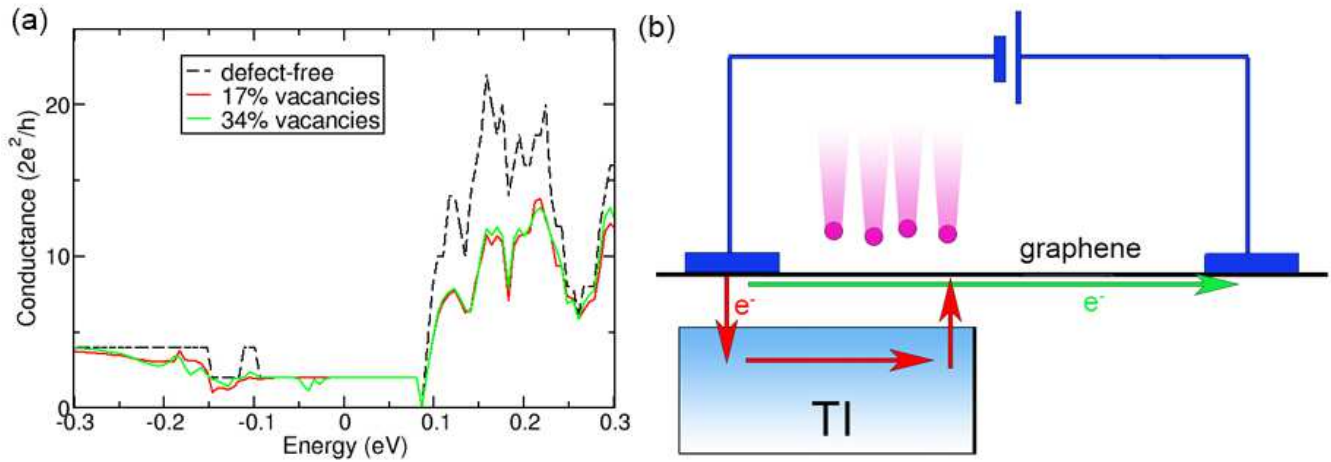}
\caption{\label{fig5} (color online) (a) Conductance of a Bi$_2$Se$_3$-contacted graphene sheet when either 17\% or 34\% C vacancies 
are introduced in graphene as compared to the conductance of a defect-free layer. (b) Schematic overview of a proposed experimental setup, 
which may prove the transfer of a topologically protected state from TI to graphene.}
\end{figure*}

As a final characterization of the hybrid state we have probed its scattering properties. In particular we have performed transport calculations, 
with the Smeagol code~\cite{sanvito-smeagol}, for the composite along the direction parallel to the graphene sheet. Note that Smeagol
provides an electronic structure for the composite essentially identical to that obtained with VASP. We have then calculated the 
system conductance for a defect-free system and for a the case where approximately 17\% or 33\% vacancies counting atoms
are introduced along the direction perpendicular to the transport [see Fig.~\ref{fig5}(a)]. Note that these are extremely large concentrations 
and here they serve the purpose to prove the topological protection of the surface state. From Fig.~\ref{fig5}(a) one can observe that
17\% of vacancies do not affect the conductance around $\ef$ indicating that the state is indeed strongly protected against back-scattering.
Even for a  33\% concentration little reduction of the conductance is found at around $\ef$, although the graphene layer is almost cut in 
two parts. Note, however, that the inclusion of impurities reduces significantly the conductance for energies away from the Fermi level,
i.e. away the topologically protected part of the graphene spectrum. This demonstrate that the transport is indeed through graphene and
it is protected against back-scattering at around the Fermi level. 

In concluding we would like to propose an experiment, which may prove the transfer of the topologically protected state to graphene. 
A schematic view of the proposed setup is shown in Fig.~\ref{fig5}(b), in which a graphene sheet is contacted only in part to 
Bi$_2$Se$_3$ while the rest remains free-standing. Defects are then introduced in graphene only at the contacting region for instance by 
electrons or ions irradiation~\cite{grapheneIrradiation}. If there exists topological protection in graphene at the contact region, the electronic 
transport through such system will not show a conductance reduction relative to that of the defect-free case. One may still argue that the 
transport through the irradiated region is via the topologically protected surface state of Bi$_2$Se$_3$ rather than through the defective 
graphene. However, in this case the charge carriers need to hop between the TI and the contacting graphene [red arrows in Fig.~\ref{fig5}(b)] 
in order to continue their flow through the bare graphene and close the electric circuit. This will degrade the conductance. In contrast if 
the transport is carried solely by states of the (defected) graphene the effects of hopping will be eliminated by the proposed geometry setup. 

Throughout the paper we have presented results as a function of the graphene/Bi$_2$Se$_3$ distance, therefore we would like to close 
this section by briefly discussing what equilibrium distance one can expect. Unfortunately this turns out to be a difficult problem. The exact
binding distance is determined by a balance between covalent and van der Waals forces. These latter ones are not captured by 
DFT local/semi-local exchange and correlation functionals and in fact we find that the two materials do 
not bind when the calculation is done at the generalized gradient approximation level~\cite{PBE}. This contrasts reality where the 
graphene/Bi$_2$Se$_3$ exists as it has been experimentally fabricated by few 
groups~\cite{Nanolett_BiSeGraphene,APL_BiSeGraphene,PRL_GrapheneSb2Te3}. 
Unfortunately the inclusion of van der Waals forces at the level of local DFT~\cite{DFTvdW} does not improve the situation, as 
screening prevents an accurate evaluation of the binding energy in layered compounds~\cite{Tilde}. However we expect the 
equilibrium graphene/Bi$_2$Se$_3$ distance to be close to the sum of Se and C covalent radii, which amounts to 2\AA. This is 
well within region A [see Fig.\ref{fig1}(c)], i.e. when the hybrid surface state forms. Besides, external pressure may be introduced 
for tuning the desired separation. 

\section{Conclusions}

In conclusion, we have investigated the electronic properties of graphene in contact to Bi$_2$Se$_3$. Three phases have been identified,
depending on the graphene/Bi$_2$Se$_3$ separation. For $d > 3.0$~\AA, the electronic structure of the composite is simply the superposition
of those of the constituents. The second phase, obtained for 2.4~\AA~$\leq d\leq$~3.0~\AA, witnesses the opening of a band-gap in graphene,
due to the annihilation of graphene states with opposite winding numbers. The third phase, when graphene and Bi$_2$Se$_3$ chemically bind, 
is the most interesting, as a topologically protected state with charge distribution mostly localized on graphene forms.\newline

\section*{Acknowledgments}

This work is sponsored by Science Foundation of Ireland (SFI) under the CSET grant underpinning CRANN and under the QDFUN project
(Grant No. 07/IN.1/I945). AN acknowledges financial support from the Irish Research Council.
Computational resources have been provided by the HEA IITAC project managed by TCHPC. \newline
\textit{Note added}: While our manuscript was under review, we became aware of a related work investigating Bi$_2$Se$_3$/graphene/Bi$_2$Se$_3$ quantum wells~\cite{kou}. Although the details of such work are different we find an agreement for the range of lattice spacing where the two sets of calculations can be compared.



\end{document}